\documentclass{ws-procs10x7}

\begin{document}

\title{Have Pentaquark States been seen?} 
\author{Volker D. Burkert}
\address{Jefferson Laboratory \\
12000 Jefferson Avenue, Newport News, VA23606, USA \\ E-mail: burkert@jlab.org}

\twocolumn[\maketitle\abstract{
The status of the search for pentaquark baryon states is reviewed in light of new results from the first two dedicated experiments from CLAS at Jefferson Lab and of new analyses from several labs on the $\Theta^+(1540)$. Evidence for and against the heavier pentaquark states, the $\Xi(1862)$ and the $\Theta_c^0(3100)$ observed at CERN and at HERA, respectively, are also discussed. I conclude that the evidence against the latter two heavier pentaquark baryons is rapidly increasing making their existence highly questionable. I also conclude that the evidence for the $\Theta^+$ state has significantly eroded with the recent CLAS results, but still leaves room for a state with an intrinsic width of  $\Gamma < 0.5$~MeV. New evidence in support of a low mass pentaquark state from various experiments will be discussed as well.}]

\section{Introduction}  The announcement in 2003 of the discovery of the $\Theta^+(1540)$, a state with flavor exotic quantum numbers and a minimum valence quark content of ($uudd\bar{s}$)\cite{leps03}, generated a tremendous amount of excitement in both the medium-energy nuclear physics and the high energy physics communities. Within less than one year the initial findings were confirmed by similar observations in nine other experiments \cite{diana,clas_d,saphir,neutrino,clas_p,hermes,zeus1,cosy,svd}, both in high energy and in lower energy measurements. These results seemed to beautifully confirm the theoretical prediction, within the chiral soliton model by D. Diakonov, M. Petrov, and M. Polyakov \cite{dpp97}, of the existence of state with strangeness $S=+1$, a narrow width, and a mass of  about 1.53 GeV. This state was predicted as the isosinglet member of  an anti-decuplet of ten states, three of which ($\Theta^+,~\Xi^{--},~\Xi^+$ ) with exotic flavor quantum numbers that experimentally can be easily distinuished from ordinary 3-quark baryons. Two observations of heavier pentaquark candidates at CERN \cite{na49} and at HERA \cite{h1} added to the expectation that a new avenue of research in hadron structure and strong QCD had been opened up.  Yet, to this day two years after the initial announcement was made, I am here to address the question if pentaquark states have really been observed. What happened?   

\section{The positive sightings of the $\Theta^+$}

A summary of the published experimental evidence for the $\Theta^+$ and the heavier pentaquark candidates is given in table \ref{tab:pentaquark_positive}. In most cases the published width is limited by the experimental resolution. The observed masses differ by up to more than 20 MeV. 
The quoted significance $S$ in some cases is based on a naive, optimistic evaluation $S = signal/ \sqrt{background}$, while for unknown background a more conservative estimate is $S = signal /\sqrt{signal + background}$, which would result in lowering the significance by one or two units. Despite this, these observations presented formidable evidence for a state at a mass of 1525-1555~GeV. A closer look at some of the positive observations begins to reveal possible discrepancies. 

\begin{table*}[t]
\caption{\small Initial positive observations of the $\Theta^+$, $\Xi_5$, and $\Theta^0_c$  pentaquark candidates.}  
\begin{tabular}{|l|c|l|c|c|}
\hline 
{Experiment} &{ Reaction} & {Energy}	& {Mass}	&  {significance}   \\
 & & (GeV) & (MeV/c$^2$) &  \\
\hline
 LEPS	& $\gamma ^{12}C \to K^- X $	& $E_{\gamma} \approx 2$ &	1540 $\pm$ 10	& 4.6$\sigma$	\\
 DIANA	& $K^+ Xe \to pK^0_s X$	& $E_{K^+} < 0.5 $	   & 1539 $\pm$ 2	&  4$\sigma$	\\  
 CLAS(d)	& $\gamma d \to pK^-K^+n$	& $E_{\gamma} < 3.8$ & 1542 $\pm$ 5	& 5.2\\
 SAPHIR	& $\gamma p \to K^0_sK^+n$	& $E_{\gamma} < 2.65$  & 1540 $\pm$ 4 $\pm$ 2	&  4.4$\sigma$	\\   
 CLAS(p)	& $\gamma p \to \pi^+K^-K^+n$	& $E_{\gamma} = 4.8 - 5.5$&1555 $\pm$ 10& 7.8$\sigma$	\\
 $\nu$BC	& $\nu A \to pK^0_s X$	& range		& 1533 $\pm$ 5		& 6.7$\sigma$	\\
 ZEUS	& $ep \to epK^0_s X$		& $\sqrt{s} = 320$	& 1522 $\pm$ 1.5 	& 4.6$\sigma$	\\
 HERMES	& $ed \to pK^0_s X$		& $E_e = 27.6$		& 1528 $\pm$ 2.6 $\pm$ 2.1& 5.2$\sigma$	\\ 
 COSY	& $pp \to \Sigma^+ pK^0_s$	& $P_p = 3$		& 1530 $\pm $5	& 3.7$\sigma$	\\
 SVD	& $pA \to pK^0_s X$		& $E_p = 70$		& 1526 $\pm 3 \pm 3$	& 5.6$\sigma$	\\	  
\hline
 NA49	& $pp \to \Xi^-\pi^- X$	& $E_p = 158$	& 1862 $\pm~2$		& 4$\sigma$		\\
\hline
 H1	& $ep \to D^{*-}p  D^{*+}\bar{p} X$ & $ \sqrt{s} = 320$ &3099	$\pm 3 \pm 5$	&  5.4$\sigma$ 	\\
\hline\hline
\end{tabular}
\label{tab:pentaquark_positive}
\end{table*}
\subsection{A problem with the width and production ratios $\Theta^+/\Lambda^*$?}
The analysis of $K^+A$ scattering data showed that the observed $\Theta^+$ state must have an intrinsically very narrow width. Two analyses of different data sets found finite width of $\Gamma = 0.9 \pm 0.3$ MeV~\cite{cahn,gibbs}, while others came up with upper limits of 1 MeV~\cite{meissner} to several MeV \cite{arndt,nussinov}. When compared with the production ratio of the $\Lambda^*(1520)$ hyperon with intrinsic  width of $\Gamma_{\Lambda^*} = 15.9$ MeV the rate of the total cross section for the formation of the two states is expected to be $R_{\Theta^+,\Lambda^*} \equiv \sigma_{tot}(\Theta^*) / \sigma_{tot}(\Lambda^*) = 0.014 $ for a 1 MeV width of the $\Theta^+$~\cite{gibbs2}. Although this relationship holds strictly for resonance formation at low energies only, dynamical models for the photoproduction of the $\Theta^+$ show that the production cross section at modest energies of a few GeV strongly depends on the width of the state~\cite{roberts04,ko04,nam05,oh04}. One therefore might expect $R_{\Theta^+,\Lambda^*}$ to remain small in the few GeV energy range. The published data however, suggested otherwise: Much larger ratios were observed than expected from the estimate based on the $\Theta^+$ width. These results, together with the upper limits obtained from experiment with null results, including the most recent results from CLAS are summarized in table \ref{tab:theta_lambda_ratio}.  

\section{Non-observations of the $\Theta^+$.}
Something else that happened in 2004 and 2005 was a wave of high energy experiments presenting high statistics data that did not confirm the existence of the $\Theta^+$ state. There are two types of experimental results, one type of experiments studies the decays of intermediate states produced in $e^+e^- $ collisions, and gives limits in terms of branching ratios. The other experiments searched for the $\Theta^+$ in fragmentation processes. Several experiments give upper limits on the $R_{\Theta^+,\Lambda^*}$ ratio. Detailed discussions of experiments that claimed sightings of a pentaquark candidate state, as well as those that generated null results are presented in detail in recent reviews~\cite{hicks05,danilov05}.  

\begin{table}[tbp]
\caption{\small The ratio $R_{\Theta^+,\Lambda^*}$ measured in various experiments. The first 5 experiments claimed a $\Theta^+$ signal, while the others give upper limits. The last two results are from the most recent CLAS measurements that give very small upper limits.}  
\begin{tabular}{|l|c|c|}
\hline 
{Experiment} & {Energy}	&  $\Theta^+/\Lambda^*$   \\
	     &   (GeV)  &   		(\%)	\\
\hline
 LEPS	& $E_{\gamma} \approx 2$ 	& $\sim 40$	\\
 CLAS(d)& $E_{\gamma} = 1.4 - 3.8$ 	& $\sim 20$	\\
 SAPHIR	& $E_{\gamma} = 1.4 - 3$  	& 10	\\   
 ZEUS	& $\sqrt{s} = 320$		& 5	\\
 HERMES	& $E_{\gamma} \approx 7 $	& $\sim 200$	\\   
\hline
CDF	& $\sqrt{s} = 1960$ 		& $<$ 3 	\\
HERA-B	& $\sqrt{s} = 42$		& $<$ 2	\\
SPHINX	& $\sqrt{s} = 12$		& $<$ 2 	\\
Belle	& $\sqrt{s} \approx 2$ 		& $<$ 2.5 	\\
BaBar	& $\sqrt{s} = 12$		& $<$ 3	\\
\hline
CLAS-2(p)	& $E_{\gamma} = 1.4 - 3.8$ 	& $<$ 0.2 \\
CLAS-2(d)	& $E_{\gamma} = 1.4 - 3.6$	&   \\
\hline
	& $K^+A$ analysis		& $\approx 1.5$  	\\
\hline\hline
\end{tabular}
\label{tab:theta_lambda_ratio}
\end{table}

\section{Are these results consistent ?} 
It is difficult to compare the low energy experiments with the high energy experiments. Low energy experiments study exclusive processes where completely defined final states are measured, and hadrons act as effective degrees of freedom. At high energies we think in terms of quark degrees of freedom, and fragmentation processes are more relevant. How can these different processses be compared quantitatively? The only invariant quantities for a resonance are quantum numbers, mass, and intrinsic width. In the absence of a resonance signal we can only place an upper limit on its width as a function of the invariant mass in which the resonance signal is expected to occur. It is therefore the total resonance width, or an upper limit on it, that allows us to compare processes for different reactions. It may not be unreasonable to assume that a narrow width of the $\Theta^+$ would result in much reduced production cross section compared to broader states such as the $\Lambda^*$ both at low and at high energies. A similar conclusion may be drawn if one considers quark fragmentation as the main source of hadron production at high energies, e.g. in $e^+e^- \rightarrow q\bar{q}$, where hadrons are generated through the creation of $q\bar{q}$ pairs from the vacuum via glue string breaking. In a scenario of independent creation of a number of $q\bar{q}$ pairs starting with a single $q\bar{q}$ pair created in $e^+e^-$ annihilation, or a single quark knocked out of a target nucleon in deep inelastic scattering, four additional $q\bar{q}$ pairs from the vaccuum are needed to form a ($uudd\bar{s}$) 5-quark object. This should be much less likely to occur than the creation of a 3-quark baryon such as the $\Lambda^*$. The latter requires creation of only two additional $q\bar{q}$ pairs. 

If we take the estimate $R_{\Theta^+,\Lambda^*} \sim 0.015$ for a 1 MeV width of the $\Theta^+$ from low energy resonance formation as a guide, we have a way of relating high energy and low energy processes.  Comparing the limits for that ratio from table \ref{tab:theta_lambda_ratio} one can make several observations: 1) The first set of experiments claiming sighting of the $\Theta^*$ show very large ratios. 2) The second set of experiments quoting upper limits are not below the ratio extracted from low energy $K^+A$ analysis. 3) The recent CLAS results are an order of magnitude below that value.  This is to be contrasted with the very large ratios measured in the first set of experiments in table \ref{tab:theta_lambda_ratio}. The focus of new experimental investigations should be to verify that these initial results are indeed correct.

\section{New results - mostly against the existence of pentaquark states.} 
During the past six months much new evidence against and some in favor of the existence of pentaquark baryons have emerged. There are new high statistics results from the CLAS detector at Jefferson Lab. New analyses of previously published data have become available from ZEUS and the SVD-2 collaborations, and LEPS studied a new channel with claimed $\Theta^+$ sensitivity. The Belle and BaBar collaborations have generated high statistics data that test the lower energy photoproduction results, and high energy experiments at Fermilab, HERA and CERN confront the claims for the $\Xi_5$ and $\Theta^0_c$ pentaquark candidates. New evidence for a doubly charged $\Theta^{++}$ comes from the STAR detector at RHIC.  These new results will be discussed in the following sections. 

\subsection{New results from CLAS.}
\begin{figure}
\epsfxsize160pt
\figurebox{120pt}{160pt}{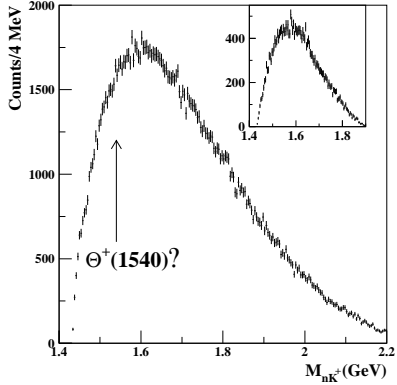}
\caption{\small\protect The invariant mass $M_{nK^+}$ from the CLAS high statistics experiment on $\gamma p \to K^0_sK^+n$.}
\label{fig:g11_nk}
\end{figure}

\begin{figure}
\epsfxsize120pt
\figurebox{120pt}{160pt}{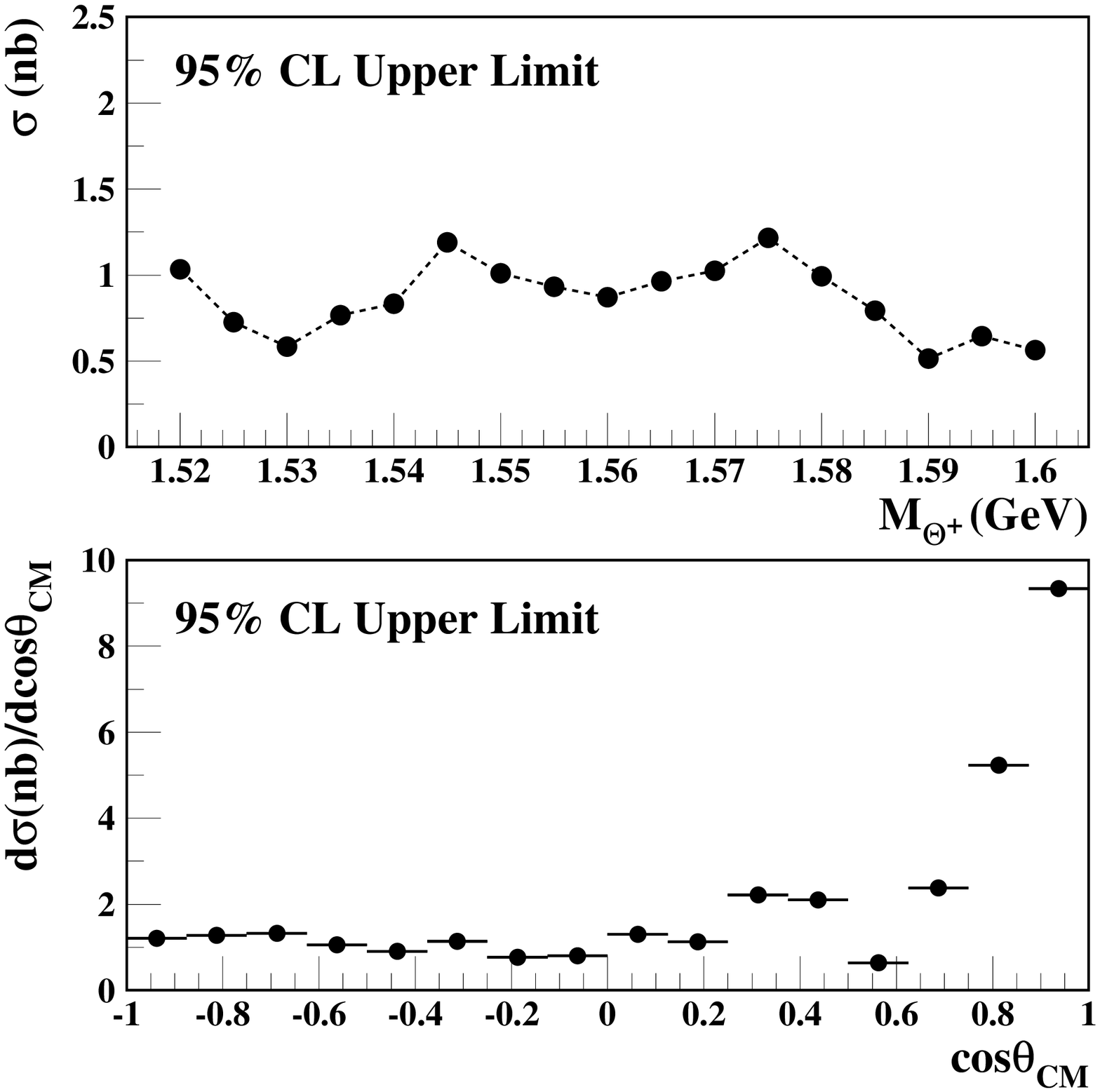}
\caption{\small\protect The CLAS upper limit on the total cross section for $\Theta^+$ production from hydrogen (top). The bottom panel shows the limit on the differential cross section at a $K^+n$ mass of 1540 MeV.}
\label{fig:g11_sigma}
\end{figure}

The CLAS collaboration has recently completed the first two dedicated high statistics experiments aimed at verifying previously reported  observations of the $\Theta^+$.  
\\
The first experiment measured the reaction $\gamma p \rightarrow K^0_sK^+(n)$, where the neutron is reconstructed using 4-momentum conservation. The $M_{K^+n}$ invariant mass distribution shown in Fig. \ref{fig:g11_nk} is structureless. An upper limit for the $\Theta^+$ cross section is derived by fitting the data with a polynomial background distribution and a sliding Gaussian that represents the experimental resolution. The upper limit at 95\% c.l. is shown in Fig.~\ref{fig:g11_sigma} versus the mass of the $\Theta^+$. In the mass range of (1.525 to 1.555)~GeV, a limit of (0.85 - 1.3)~nb (95\% c.l.) is derived. What does this result tell us? 
There are several conclusions that can be drawn from the CLAS result on the proton. \\
1) It directly contradicts, by two orders of magnitude in cross section, the SAPHIR  experiment\cite{saphir} that claimed a significant signal in the same channel and in the same energy range, and published a cross section of 300nb for $\Theta^+$ production. \\
2) Together with the extracted $\Lambda^*$ cross section it puts an upper limit on the $\Theta^+/\Lambda^*$ ratio in table \ref{tab:theta_lambda_ratio} that is an order of magnitude lower than the value from the $K^+A$ analysis, and strongly contradicts the 'positive' $\Theta^+$. 
\\
3) It puts a very stringent limit on a possible production mechanism. For example, it implies a very small coupling $\Theta^+NK^*$ which in many hadronic models was identified as a major source for $\Theta^+$ production. \\     
4) If there is no large isospin asymmetry in the elementary process, the $\gamma D$ and $\gamma A$ experiments at lower statistics should not be able to see a signal. Possible mechanisms to obtain a large isospin asymmetry have been discussed in the literature following the first announcement of the new CLAS data\cite{nam05_2,lipkin05}.   
\\
\begin{figure}[t]
\epsfxsize150pt
\figurebox{120pt}{160pt}{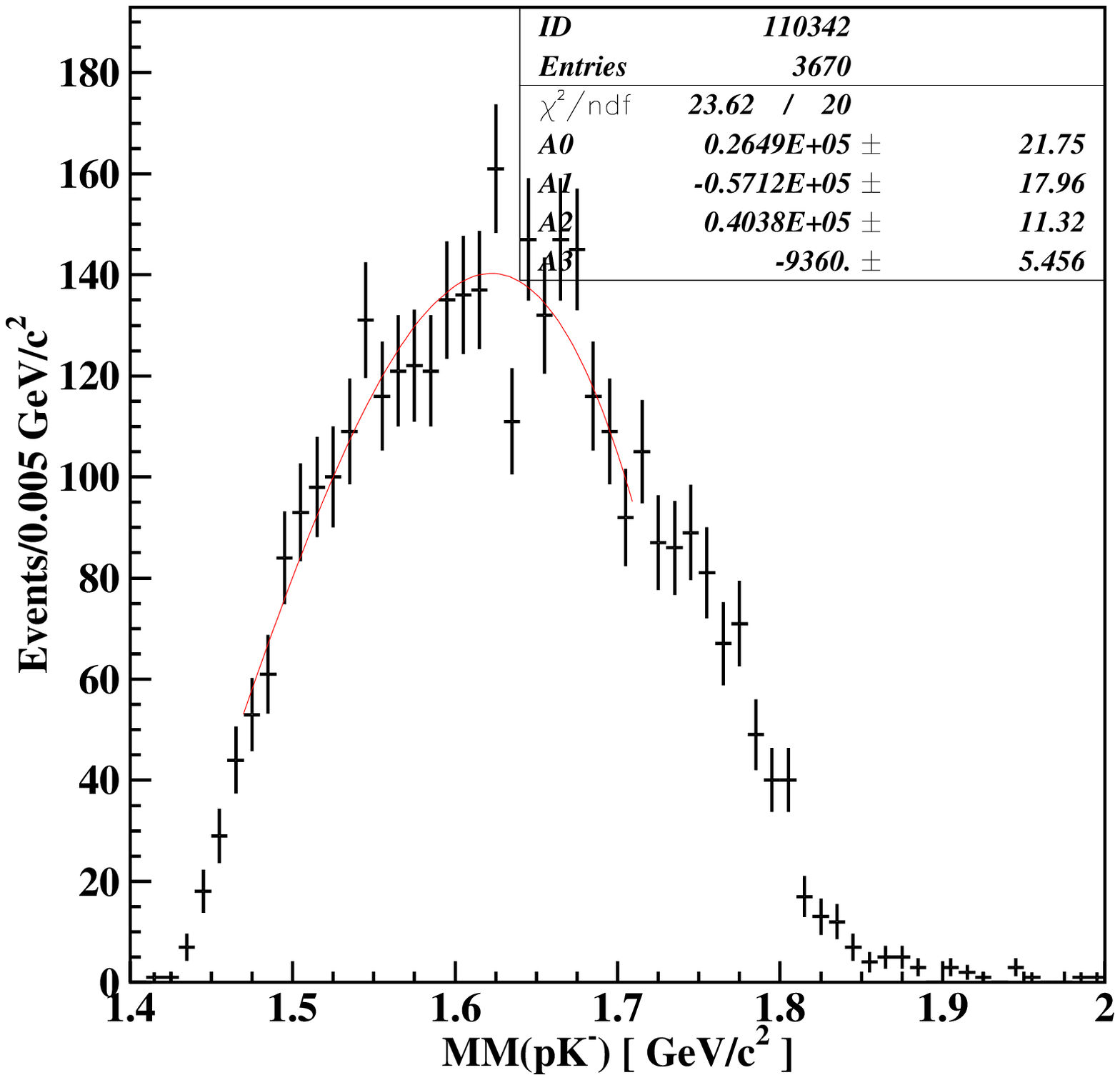}
\caption{\small\protect The missing mass $M_{pK^-}$ from the CLAS high statistics deuterium experiment. Events are selected with the same kinematics as the previously published results.}
\epsfxsize160pt
\figurebox{120pt}{160pt}{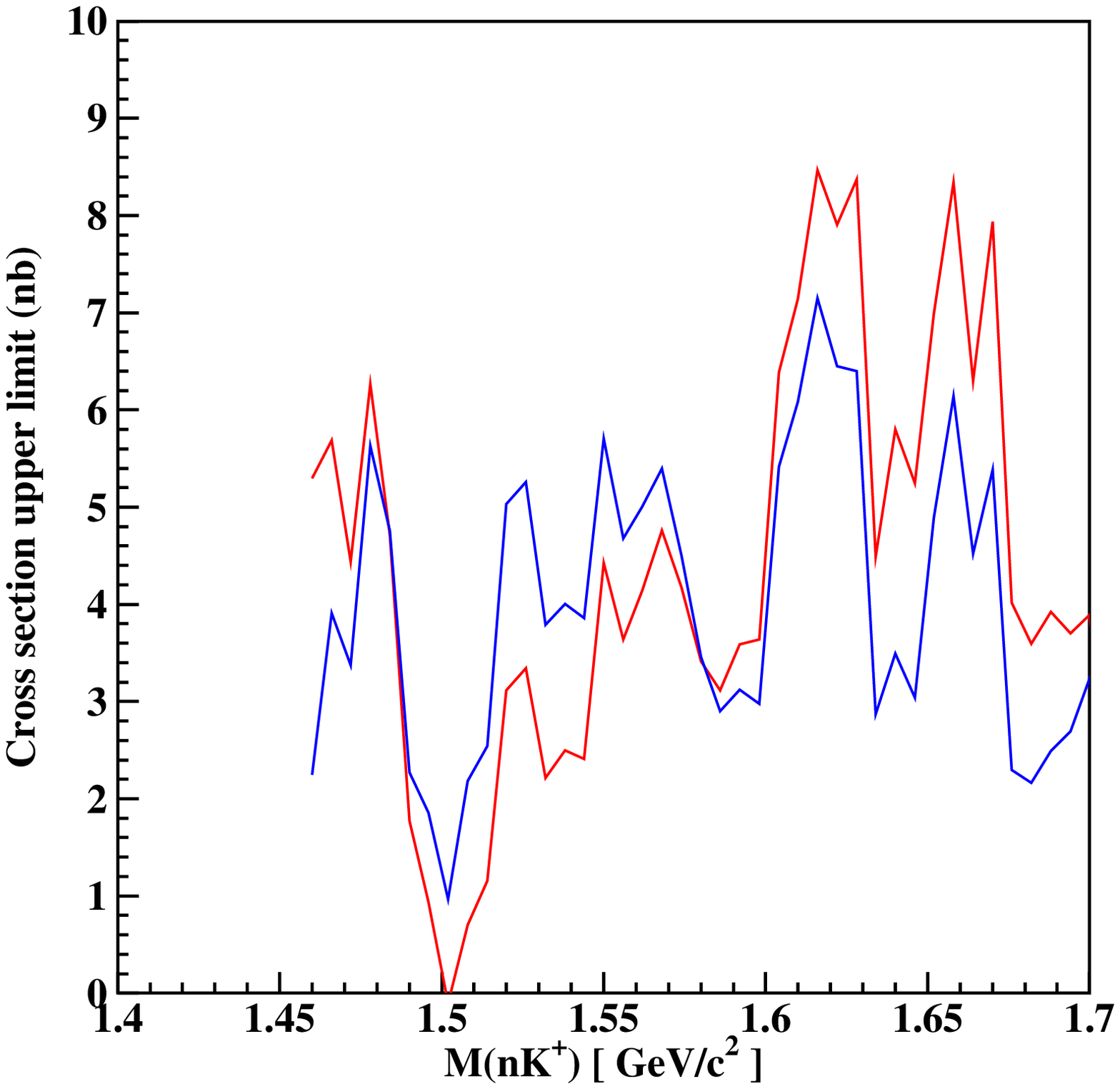}
\caption{\small\protect  The upper limit (95\% c.l.) of the total cross section for $\Theta^+$ photoproduction from neutrons.}
\label{fig:g10_nk}
\end{figure}
The second new CLAS experiment measured the reaction $\gamma D \rightarrow pK^-K^+(n)$, where the neutron again is reconstructed from the overdetermined kinematics. This experiment represents a dedicated measurements to verify a previous CLAS result that claimed more than 4.6$\sigma$ significance for the $\Theta^+$ in the same channel and same energy range. The aim is to measure the possibly preferred production on neutrons through $\gamma n \rightarrow K^-K^+n$ . To avoid the complication of precise neutron detection the recoil proton is measured instead, requiring momenta of greater than 0.3 GeV/c for the proton to be detected in CLAS. This reduces the acceptance for the exclusive reaction by a large factor. The preliminary results, representing about 50\% of the full statistics, are shown in Fig. \ref{fig:g10_nk}. Again, no significant signal is seen in a data sample with about seven times the statistics of the previous result.  From this result an upper limit of 5nb (95\% c.l.) is  derived for the elementary cross section on the neutron. The limit is somewhat model-dependent as rescattering effects in the deuteron must be taken into account. The result clearly contradicts the previous lower statistics data. In order to understand the discrepancy the older data have been reanalyzed with a background distribution extracted from the new high statistics data set. The results show an underestimation of the background normalization in the original analysis. A new fit with the improved background yields a signal with a significance of  3$\sigma$ compared to the $(5.2 \pm 0.6)\sigma$ published. 

What is the impact of the combined CLAS data on proton and neutron? For this we compare the cross section limits with various dynamical model calculations~\cite{roberts04,ko04,nam05,oh04}. In hadro-dynamical models, the cross section is computed based on an effective Lagrangian approach. The comparison is shown in table \ref{tab:dynamical_models} for the $J^P = 1/2^+$ assignment and $\Gamma_{\Theta^+} = 1$~MeV. The upper limit for the combined proton and neutron targets would be less than 0.5 MeV for at least one of the targets in each model. 

\begin{table}[tbp]
\caption{\small Limit on $\Theta^+$ width from recent CLAS results. Upper limits for the total cross section on protons of 1.25nb, and on neutrons of 4nb are used to determine the limit on the width. The first line in each row is for $\gamma p$, the second line is for $\gamma n$. The cross section is computed for a $J^P = 1/2^+$ assignment of the $\Theta^+$ and a width of 1~MeV.}  
\begin{tabular}{|l|c|c|}
\hline 
{Publication} 	& $\sigma(\gamma N)$ 	&  $\Gamma_{\Theta^+}$  \\
 		& (nbarn)			&   (MeV)		\\
\hline
S. Nam et al.  \cite{nam05}	& 2.7			& $< 0.5$ 		\\
			& 2.7			& $< 1.7$ 		\\
\hline
Y. Oh et al., \cite{oh04}	& $\sim 1.6$ 		& $< 0.8$ 	\\	
			& $\sim 8.7$			& $< 0.5$ 	\\
\hline 
C.M Ko et al., \cite{ko04}	& 15				& $< 0.08$	\\
			& 15				& $< 0.25$	\\
\hline
W. Roberts \cite{roberts04}	& 5.2				& $< 0.24$	\\
			& 11.2				& $< 0.4 $	\\	  
\hline\hline
\end{tabular}
\label{tab:dynamical_models}
\end{table}
\subsection{BaBar study of quasi-real photoproduction}

\begin{figure}[t]
\epsfxsize180pt
\figurebox{120pt}{20pt}{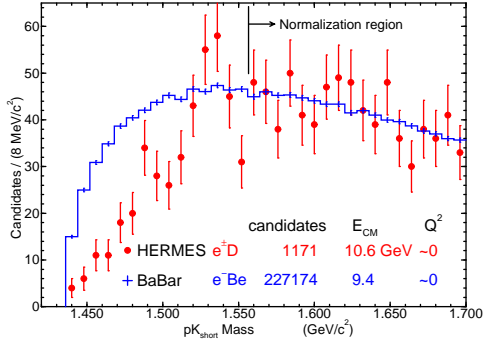}
\caption{\small\protect Comparison of BaBar results on quasi-real photoproduction of $pK^0_s$ from beryllium, with the HERMES results. The very high statistics BaBar data do not show any structure near 1530 MeV, while the HERMES data do.  The falloff of the HERMES data near the lower mass end may indicate acceptance limitations. }
\label{fig:babar_hermes}
\end{figure}

The BaBar collaboration also studied the quasi-real photoproduction of $e+Be \rightarrow p K_s^0 + X$~\cite{babar_hermes}. In this case electrons with energies of $\sim 9$~GeV resulting from small angle scattering off the positron beam interact with the beryllium beam pipe. The scattered electron is not detected, and the invariant mass of final state inclusive $pK^0_s$  is studied for possible contributions from $\Theta^+ \rightarrow pK^0_s$.  There is no evidence for a signal. The data can be directly compared to the HERMES results which were taken in quasi-real photoproduction kinematics from deuterium at higher electron beam energies. The comparison is shown in Fig.~\ref{fig:babar_hermes}. While at high masses the two distributions coincide, a potential loss of acceptance at HERMES is seen for low mass $pK^0_s$ pairs. The HERMES peak may, at least in part, be the result of the acceptance rising up below the nominal $\Theta^+$ mass. The absence of any signal in the high statistics BaBar data calls the signal observed by HERMES into question. BaBar also compare their null results with the ZEUS signal observed at $Q^2 > 20$~GeV$^2$. However, since ZEUS sees no signal at low $Q^2$ and BaBar only probes the quasi-real photoproduction kinematics, this comparison is indeed misleading.        
\begin{figure}[t]
\epsfxsize180pt
\figurebox{120pt}{160pt}{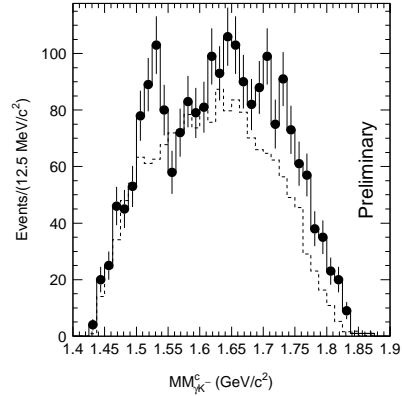}
\caption{\small\protect  The LEPS data on deuterium. The missing mass distribution $MM({\gamma K^-})$ showing a peak at 1530 MeV. }
\label{fig:leps2}
\end{figure}
\subsection{LEPS at SPring-8}
 
\begin{figure}[tbp]
\epsfxsize180pt
\figurebox{120pt}{160pt}{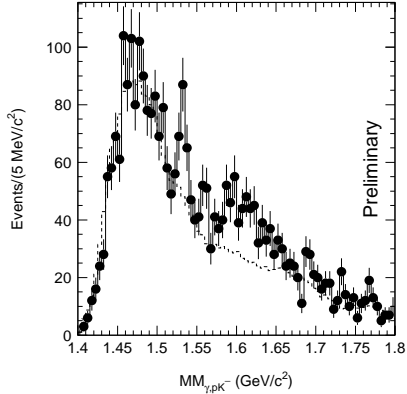}
\caption{\small\protect  The LEPS data on deuterium.The missing mass distribution $MM_{\gamma, K^-p}$ with events selected with the invariant mass $M_{K^-p}$ near the $\Lambda^*$. A peak is seen at a mass of 1530 MeV, and is interpreted as the $\Theta^+$. }
\label{fig:leps_lambda*}
\end{figure}

The LEPS experiment originally claimed the discovery of  the  $\Theta^+$ in photoproduction from a carbon target in the inclusive reaction $\gamma C \rightarrow K^- X$ plotting the Fermi-momentum corrected missing mass $M_X$. The experiment has been repeated with a liquid deuterium target and higher statistics. The still preliminary results are shown in Fig.~\ref{fig:leps2}. A peak at 1530 MeV is observed. The data also show a large ratio of $\Theta^+/\Lambda^*$ (see table \ref{tab:theta_lambda_ratio}). Since these data are obtained at energies similar to the new CLAS data on deuterium, they need to be confronted with the recent exclusive CLAS data taken on deuterium in the reaction $\gamma d \rightarrow K^- p K^+ n$, and the resulting cross section limit for the elementary cross section on neutrons. This will require extraction of a normalized cross section from the LEPS data. There are also new results from LEPS on the channel $\gamma d \rightarrow \Lambda^* X$. The mass distribution $M_{K^+n}$ is shown in Fig.~\ref{fig:leps_lambda*}. After accounting for background from $\Lambda^*(1520)$ and sideband subtraction, a narrow peak near 1530~MeV/c$^2$ with  5$\sigma$ significance is claimed. The signal emerges only when events are selected with $M_{K^-p} \sim M_{\Lambda^*}$, indicating that the process $\gamma D \rightarrow \Lambda^* \Theta^+$ may be observed.  The mass distribution also shows an excess of events near 1600 MeV/c$^2$.     


\subsection{Results from {Belle}}

\begin{figure}[t]
\epsfxsize160pt
\figurebox{120pt}{160pt}{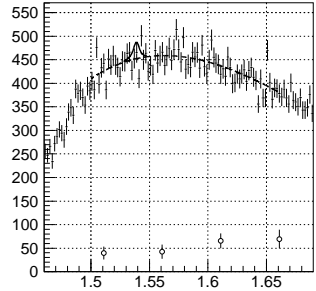}
\caption{\small\protect The Belle $pK^0_s$ mass spectrum.  }
\label{fig:belle_mass}
\end{figure}

\begin{figure}[tbp]
\epsfxsize160pt
\figurebox{120pt}{160pt}{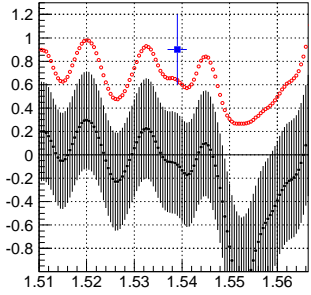}
\caption{\small\protect The Belle upper limit on the $\Theta^+$ width. The dotted line shows the 90\% c.l. limit for the width. The data point is result of the analysis of the DIANA result in $K^+ Xe \to K^0_s p + X$.}
\label{fig:belle_width}
\end{figure}

New results by the Belle collaboration have been presented recently\cite{belle}. Belle uses hadrons created in high energy $e^+e^-$ collisions and reconstructs the hadron interaction with the vertex detector materials. The momentum spectrum is sufficiently low so that resonance formation processes such as $K^+n \rightarrow \Theta^+ \rightarrow pK^0_s$ can be studied. The high statistics $pK^0_s$ invariant mass shows no signal, as is seen in Fig.~\ref{fig:belle_mass}. The upper limit on the formation cross section can be used to extract an upper limit for the $\Theta^+$ width, which is shown in Fig. \ref{fig:belle_width}. At a specific mass of 1539 MeV, an upper limit of $\Gamma_{\Theta^+} < 0.64$ MeV (90\% c.l.) is derived. The mass corresponds to the $\Theta^+$ mass claimed by the DIANA experiment~\cite{diana}. However, if one allows the entire mass range for the $\Theta^+$  from 1525 to 1555 MeV claimed by experiments, the upper limit would be $\Gamma_{\Theta^+} < 1 $ MeV (90\% c.l.). The latter value confirms the limit derived in previous analyses.   

\subsection{BaBar results in quark fragmentation}  

\begin{figure}[t]
\epsfxsize180pt
\figurebox{120pt}{160pt}{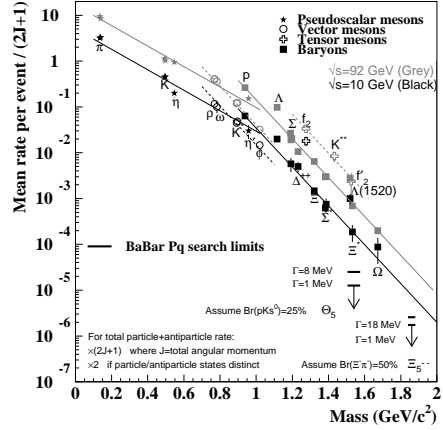}
\caption{\small\protect The BaBar baryon mass spectrum.  }
\label{fig:babar_baryons}
\end{figure}

The BaBar collaboration at SLAC searches for the $\Theta^+$ as well as the $\Xi^{--}$ pentaquark states directly in $e^+e^-$ collisions~\cite{babar_hadrons}, mostly in the  quark fragmentation region. With high statistics no signal is found for either $\Theta^+(1540)$ or $\Xi^{--}(1862)$, and upper limits are placed on their respective yields. The results are shown in Fig.~\ref{fig:babar_baryons}.  The limit on the production rates are 8 or 4 times lower than the rates of ordinary baryons at the respective masses. It is, however, not obvious what this result implies. The slope  for the production of pseudoscalar mesons is $d(event~rate)/d(mass) = 10^{-2}$/GeV. For 3-quark baryons it is $10^{-4}$/GeV, i.e. the rate drops by a factor of 10,000 per one unit of GeV in mass.  In the quark fragmentation region, if we extrapolate from mesons where only one $q\bar{q}$ pair must be created to form a meson starting with one of the initial quarks in the $e^+e^-$ annihilation, and baryons where two $q\bar{q}$ pairs are needed, to pentaquarks where four $q\bar{q}$ pairs are needed, the slope for pentaquark production in fragmentation would be $10^{-8}$/GeV. Since there is no rate measured for a pentaquark state there is no normalization point available. If we arbitarily normalize the pentaquark line at the point where baryon and meson lines intersect, the line falls one order of magnitude below the upper limit for the $\Theta^+$ and several orders of magnitude below the upper limit for the $\Xi^{--}$ assuming a mass of 1862 MeV for the latter. The sensitivity of quark fragmentation to 5-quark baryon states is thus questionable. Moreover, the limit for $R_{\Theta^+,\Lambda^*} < 0.02$ at 95\% c.l. is not in contradiction with the ratio estimated at low energy assuming a width of $\Gamma_{\Theta^+} = 1$ MeV~\cite{gibbs}.  
 
\subsection{New $\Theta^+$ analysis from ZEUS}

\begin{figure}[tbp]
\epsfxsize160pt
\figurebox{120pt}{160pt}{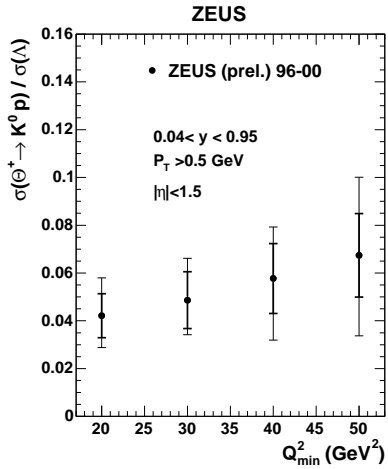}
\caption{\small\protect  $Q^2$-dependence of the $\Theta^+/\Lambda^*$ ratio measured by ZEUS.}
\label{fig:zeus_theta_lambda}
\end{figure}

The ZEUS collaboration has extended the analysis of their $\Theta^+$ signal and studied possible production mechanisms\cite{zeus2}. The signal emerges at $Q^2 > 20$~GeV$^2$ and remains visible at $Q^2 > 50$~GeV$^2$.  The $\Theta^+$ and $\bar{\Theta}^-$ signals are nearly equally strong, however, the signal is present only at forward rapidity $\eta^{lab} > 0 $ and not visible at backward rapidity $ \eta^{lab} < 0 $. There is currently no possible production mechanism that would generate such a pattern. The ZEUS collaboration extracted the $Q^2$-dependence of the ratio $R_{\Theta^+, \Lambda^*}$ which is shown in Fig.~\ref{fig:zeus_theta_lambda}. It shows a weak dependence on $Q^2$ .  

\subsection{New results from high energy hadronic interaction experiments}

\begin{figure}[t]
\epsfxsize140pt
\figurebox{120pt}{160pt}{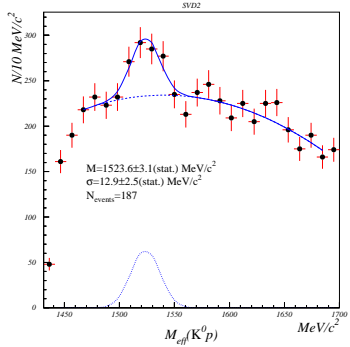}
\caption{\small\protect  Results of a new analysis by the SVD-2 group of their published data.}
\label{fig:svd2}
\end{figure}

The SVD-2 collaboration has reanalyzed their published data\cite{svd2} with much improved event reconstruction efficiency. The experiment measured the reaction $pA \rightarrow pK^0_s X$ using a 70 GeV incident proton beam. The main component in the detector system is the silicon vertex detector (SVD). Events are divided into two samples: events with the $K^0_s$ decaying inside and events with the decay outside the SVD. The two distributions both show a significant peak at the mass of ~1523 MeV/c$^2$. One of the distributions is shown in Fig.~\ref{fig:svd2}. A combined significance for two independent data sets of $\sim7.5~\sigma$ is obtained. The strangeness assignment in the $pK^0_s$ channel is not unique, and could also indicate excitation of a $\Sigma^*$ resonance. In this case one would expect a decay $\Sigma \rightarrow \Lambda \pi$, which is not observed. Therefore, an exotic $S = +1$ assignment of that peak is likely should it be a resonant state.
\begin{figure}[tbp]
\epsfxsize140pt
\figurebox{120pt}{160pt}{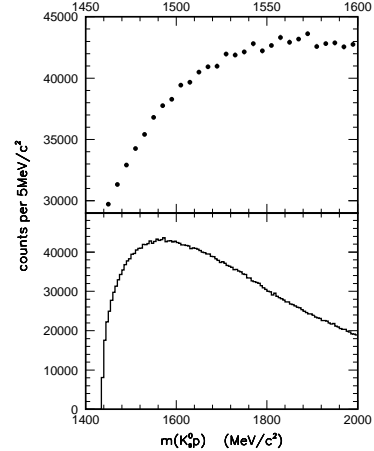}
\caption{\small\protect Invariant mass $M_{K^0p}$ measured in the CERN WA89 hyperon experiment.}
\label{fig:wa89}
\end{figure}
The SVD-2 results have been challenged by the WA89 collaboration that measured the process $\Sigma^- A \to pK^0_s X$ in comparable kinematics~\cite{wa89}. Their  mass distribution, shown in Fig.~\ref{fig:wa89}, does not exhibit any signal in the mass range of the $\Theta^+$ candidate. The WA89 collaboration claims that their results are incompatible with the SVD results.    
\begin{figure}[t]
\epsfxsize200pt
\figurebox{120pt}{160pt}{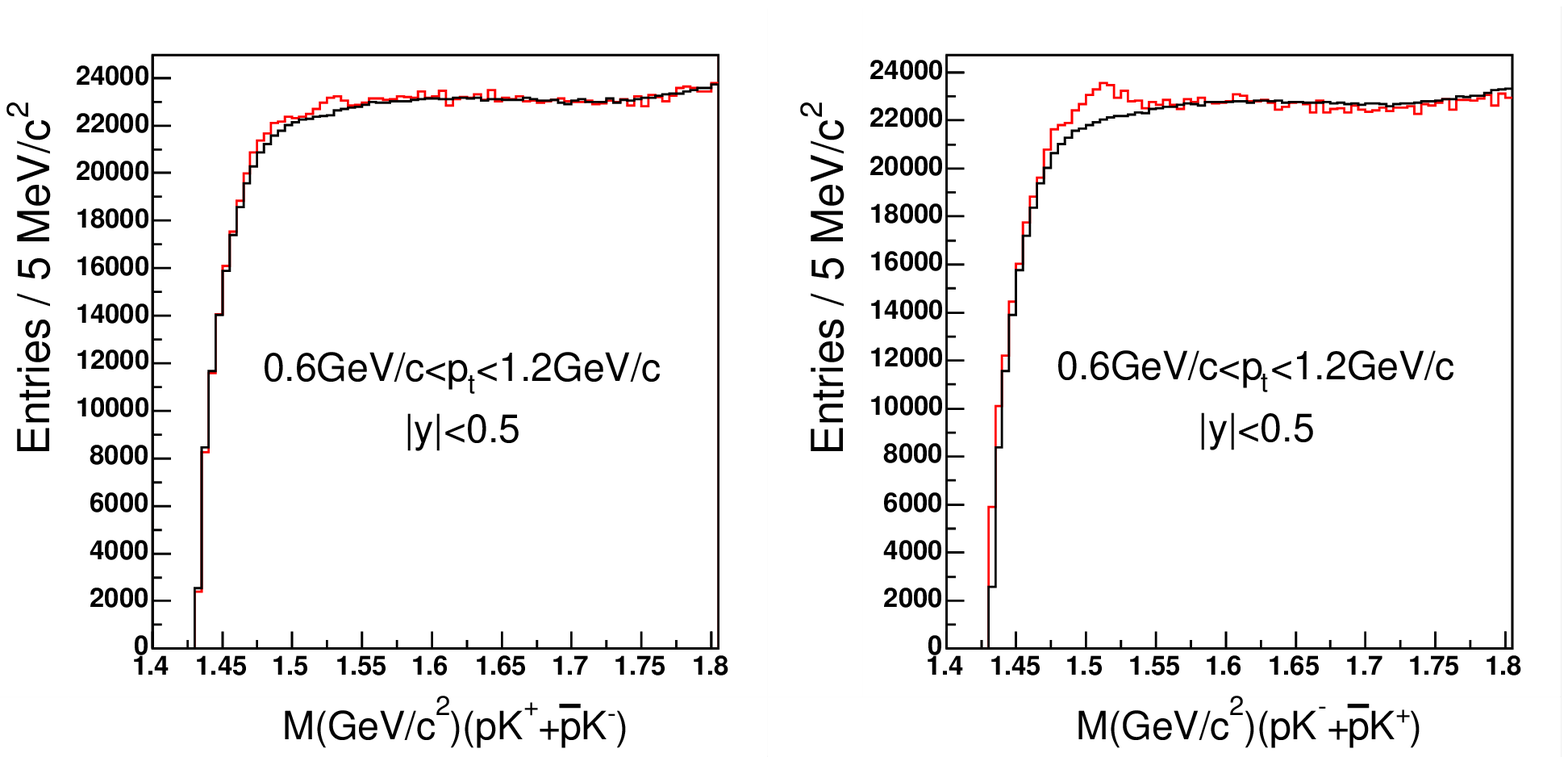 }
\epsfxsize200pt
\figurebox{120pt}{160pt}{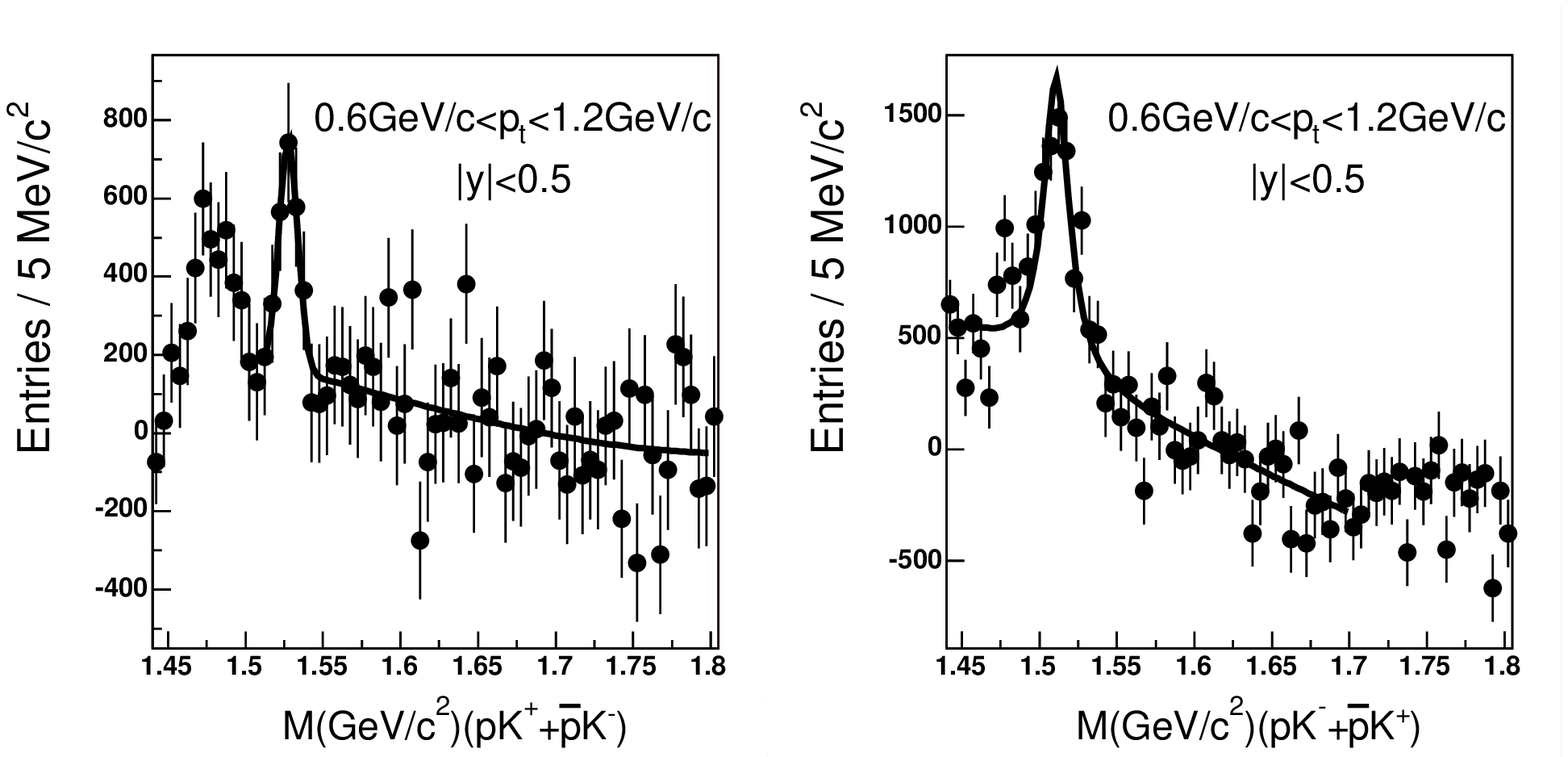}
\caption{\small\protect Invariant mass $pK^+ + \bar{p}K^-$ (left) and $K^-p + K^+\bar{p}$ (right) measured by STAR. The spectra are shown before (top) and  after (bottom) subtraction of background from mixed events. The l.h.s. shows the $\Theta^{++}$ candidate signal, the r.h.s shows the $\Lambda^*$ signal.}
\label{fig:star}
\end{figure}
\section{An isovector $\Theta^{++}$ candidate?}

Inspired by the prediction of Diakonov et al., of an anti-decuplet of 5-quark states, with the $\Theta^+$ being an isoscalar, the focus of the search for the lowest mass pentaquark was on an isoscalar baryon with $S=+1$. However, searches have also been conducted for a possible isovector baryon state with charge $Q=+2$. The final state to study is $pK^+$. No signal was seen in  any of these searches. However, recently the STAR collaboration at RHIC presented data indicating a small but significant $\Theta^{++}$ candidate~\cite{huang}. The data are shown in Fig.~\ref{fig:star} before and after background subtraction. A peak with a significance of $5\sigma$ is seen at a mass of about 1530 MeV/c$^2$ in the d-Au collision sample. The $\Lambda^*$ signal is also clearly visible.

If the $\Theta^{++}$ signal is real, then there must be also a signal in the singly charged channel, i.e. a $\Theta^{+}$ . A small peak with relatively low significance appears in the $K_s^0p$ invariant mass spectrum, however shifted by about 10 MeV/c$^2$ to higher mass values.   

\begin{figure}[t]
\epsfxsize200pt
\figurebox{120pt}{160pt}{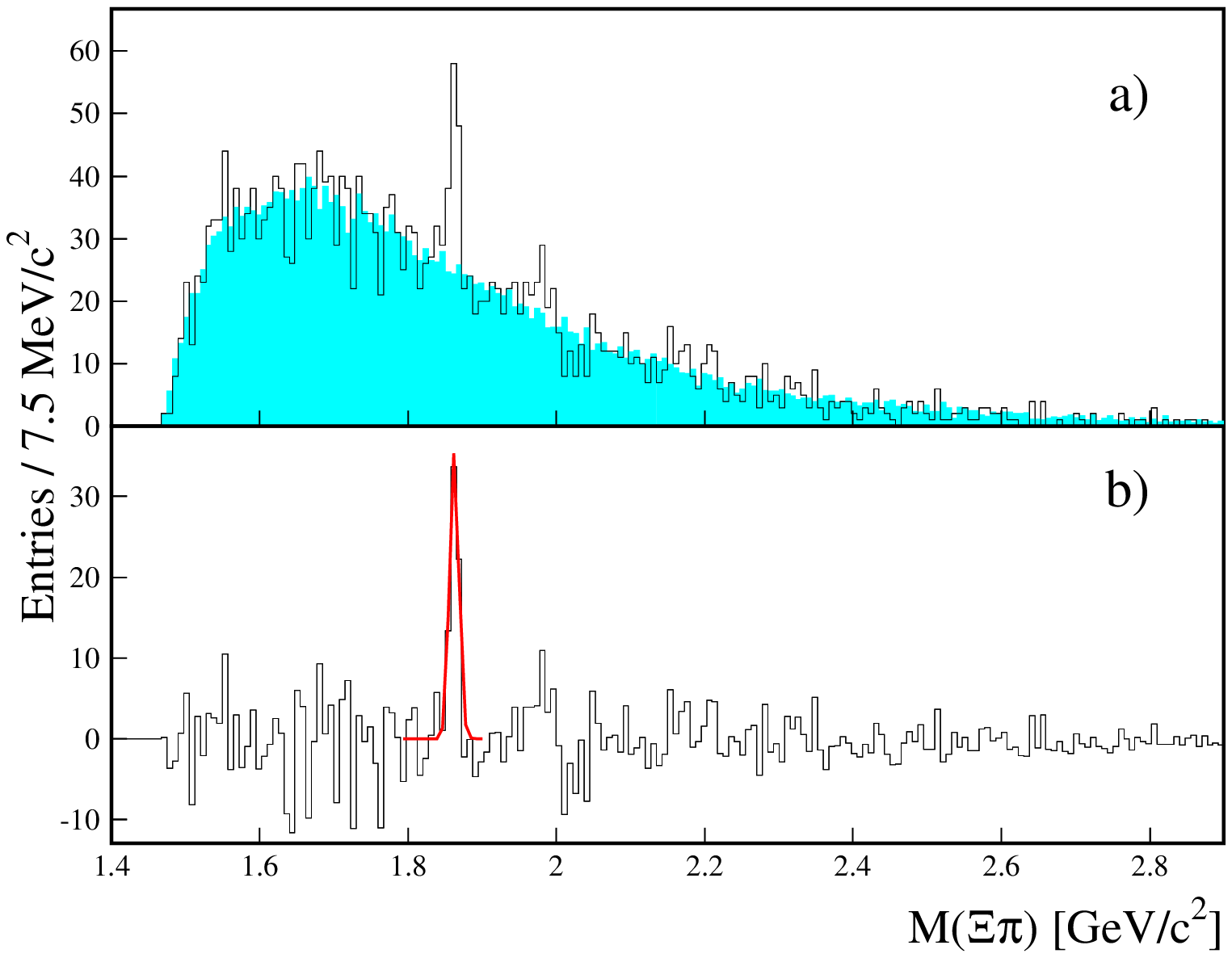}
\epsfxsize180pt
\figurebox{120pt}{160pt}{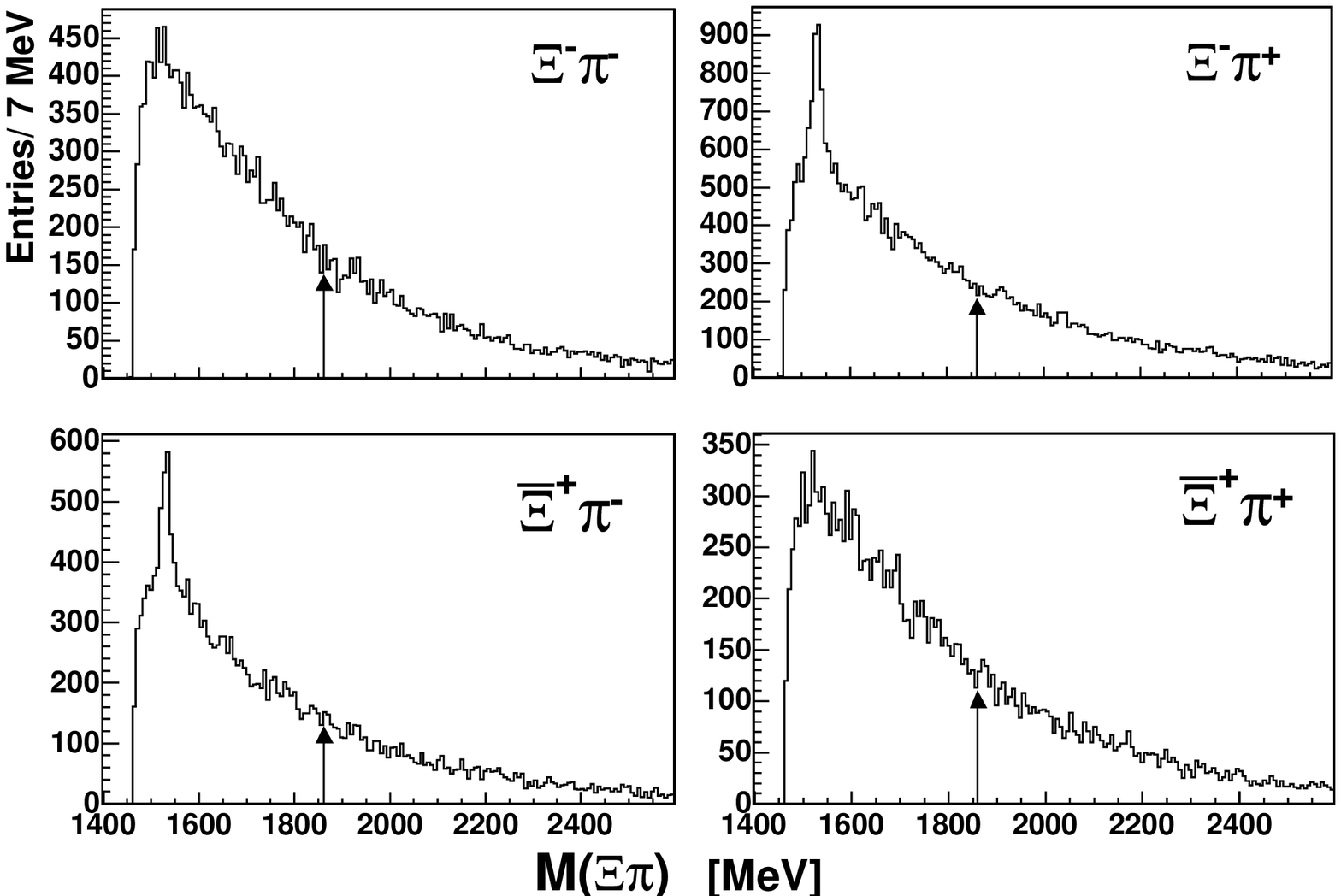}
\caption{\small\protect  The NA49 results (top) on the claimed $\Xi_5(1862)$. The sum of the $\Xi^-\pi^-$ and $\Xi^-\pi^+$ distributions are shown before and after background subtraction. The COMPASS results are displayed in the bottom panel. The $\Xi(1530)$ state is clearly seen in the neutral charge combinations, while the $\Xi_5$ expected at 1862 MeV is absent.}
\label{fig:cascade}
\end{figure}

So far I have focussed in my talk on the $\Theta^+$, as without the evidence for the $\Theta^+$ there would not have been any search for other pentaquark states within the anti-decuplet. However, much effort has been put recently into the search for the two heavier pentaquark candidates claimed in two high energy experiments.     

\section{Status of  $\Xi_5$ and $\Theta_c^0$}

\begin{figure}[t]
\epsfxsize200pt
\figurebox{120pt}{160pt}{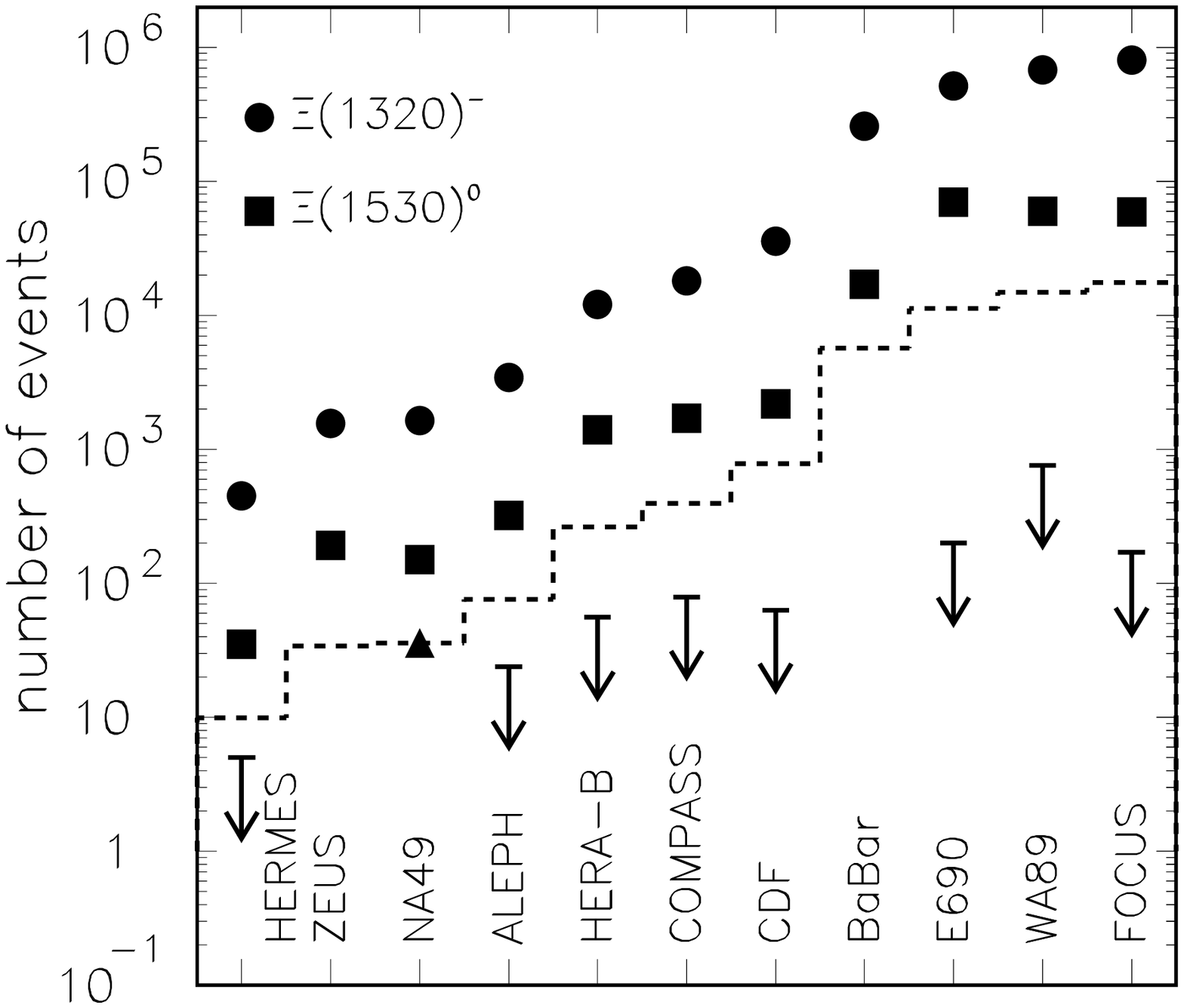}
\caption{\small\protect Summary of experimental results in the search for the exotic cascade $\Xi_5$. A comparison of the number of observed ground state $\Xi^-(1320)$ and excited state $\Xi^0(1530)$ is shown. The arrows indicate upper limits for the number of exotic $\Xi_5$ candidates. Only NA49 has observed a signal.}
\label{fig:cascade_summary}
\end{figure}

A candidate for a 5-quark $\Xi_5$ has been observed in the $\Xi^-\pi^-$ final state by the CERN NA49 experiment, and a candidate $\Theta^{0}_c$ for the charmed equivalent of the $\Theta^+$ has been claimed by the H1 experiment at HERA in the channel $D^*p$. In contrast to the $\Theta^+$, which has been claimed in  at least ten experiments, the heavier candidates have not been seen in any other experiment. The $\Xi_5$ state of NA49 has been searched for by several experiments\cite{babar_cascade,cdf_cascade,focus_cascade,hera-b_cascade,hermes_cascade,wa89_cascade,zeus_cascade,compass_cascade}. The ratios $\Xi_5/\Xi(1530)$ determined by several experiments are shown in table \ref{tab:cascade_limits}. However, the highest energy experiments probe production through quark fragmentation, and may not be directly comparable to the NA49 results. The FOCUS photoproduction experiment and the COMPASS muon scattering experiment are close to the kinematics of NA49. The COMPASS results are shown in Fig.~\ref{fig:cascade} and compared to the original NA49 results. No signal is observed. FOCUS also did not observe a signal and obtained an upper limit nearly two orders lower than the signal seen by NA49.
\begin{table*}[t] 
\begin{centering}
\caption{\footnotesize Results of searches for the $\Xi_5$ pentaquark state.}
\begin{tabular}{|l|c|c|l|}
\hline 
 Experiment 	& Initial state 	& Energy (GeV)		&  $\Xi_5^{--}/\Xi(1530)$  \\
\hline\hline 
 NA49  		&  pp		&  $E_p = 158$		&  $0.24$	 \\
\hline
 COMPASS	& $\mu^+ A$	&  $E_{\mu} = 160$	&  $< 0.046$	\\
 ALEPH		&  $e^+e^-$	&  $\sqrt{s} = M_Z$	&  $< 0.075$  	\\
 BaBar		&  $e^+e^-$	&  $m_{Y(4s)}$		&  $< 0.0055$	\\
 CDF		&  $p\bar{p}$	&  $\sqrt{s} = 1960$	&  $< 0.03$	\\
 E690 		&  $pp$		&  $E_p = 800$		&  $< 0.003$	\\
 FOCUS		&  $\gamma p$	&  $E_{\gamma} < 300$	&  $< 0.003$	\\
 HERA-B		&  $pA$		&  $E_p = 920$		&  $< 0.04$	\\
 HERMES		&  $eD$		&  $E_e=27.6$		&  $< 0.15$	\\
 WA89		&  $\Sigma^-A$	&  $E_{\Sigma}= 340$	&  $< 0.013$	\\
 ZEUS		&  $ep$		&  $\sqrt{s} = 310$	&  not seen	\\
\hline\hline
\end{tabular}
\end{centering}
\label{tab:cascade_limits}
\end{table*}
A summary of the  search for the $\Xi_5$ taken from the COMPASS paper~\cite{compass_cascade} is shown in Fig.~\ref{fig:cascade_summary}.
 
\begin{figure}[t]
\epsfxsize200pt
\figurebox{120pt}{160pt}{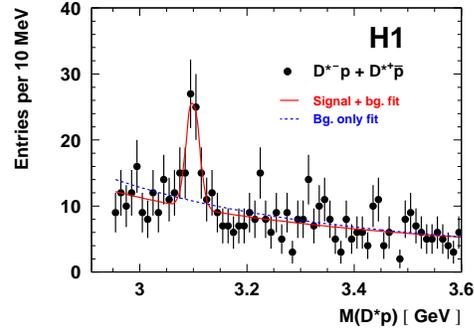}
\epsfxsize180pt
\figurebox{120pt}{160pt}{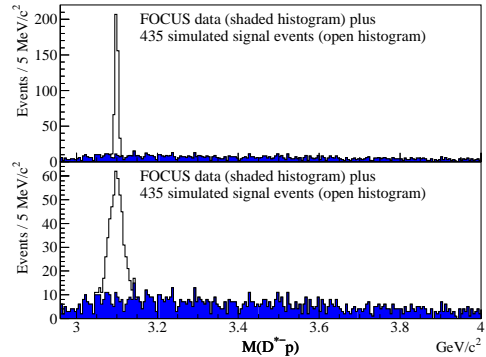}
\caption{\small\protect  Top: The results of H1 on a charmed $\Theta^0_c(3100)$ candidate. Bottom: Results of searches on the $\Theta^0_c$ candidate by FOCUS. The colored area shows the measured distributions, while the solid line indicates the expected signal extrapolated from the H1 measurement.}
\label{fig:focus_charmed}
\end{figure}

The $\Theta^0_c(3100)$ pentaquark candidate so far has only been seen by the H1 experiment at HERA. Several other experiments came up empty-handed~\cite{negative_charmed_theta},  and ZEUS and FOCUS claim incompatibility of their results with the H1 findings. The H1 results and the FOCUS results are shown in  Fig.~\ref{fig:focus_charmed}.


\section{Summary and conclusions}

Over the past year the evidence for the existence of pentaquark baryons has clearly lost much of its original significance. 

The evidence for the two heavy pentaquark candidate states, the $\Xi_5(1862)$ and the $\Theta^0_c(3100)$, observed at unexpected masses, and each seen in one experiment only, has been drastically diminished. Several experiments with high sensitivity to the relevant processes have found no indication of these states. In the face of overwhelming evidence against these states, experiments claiming positive sightings should either explain why other experiments are not sensitive, or should re-evaluate their own results. 

The situation with the $\Theta^+$ state observed at masses near 1540 MeV is less clear, although evidence for the state has also diminished significantly. So far more than ten experiments claimed to have obverved a narrow state with exotic flavor quantum number $S=+1$. Two of the initial results (SAPHIR and CLAS(d)) have been superseded by higher statistics measurements from CLAS~\cite{g11,g10} conducted at same energies and with same or overlapping acceptances. No signal was found in either case. In addition, the HERMES results are being challenged by new high statistics data from BaBar~\cite{babar_hermes}. It is remarkable that experiments claiming a $\Theta^+$ signal, measured $\Theta^+ / \Lambda^*$ ratios much above values naively expected from $K^+ A$ scattering analysis (see table \ref{tab:theta_lambda_ratio}).  HERMES even measures a $\Theta^+$ cross section that is significantly higher than the cross section for $\Lambda^*$ production.  

The Belle experiment~\cite{belle} studying $K^+A$ scattering, is beginning to challenge the DIANA results, the second experiment claiming observation of the $\Theta^+$. Belle extracted an upper limit of 0.64 MeV (90\% c.l.) for the width of any $\Theta^+$  signal at a mass of 1539 MeV. In the larger mass range of 1525-1555 an upper limit of 1 MeV has been extracted. This is to be contrasted with the width extracted from the DIANA experiment as well as from $K^+D$ scattering of $\Gamma_{\Theta^+} = 0.9 \pm 0.3$ MeV. However, the Belle limit is not (yet) in strict contradiction to the DIANA results at this point. It will be interesting to see if the Belle limit can be further reduced with higher statistics.  

The new CLAS results on protons and neutrons also challenge the value of the $\Theta^+$ width. Using hadronic models, upper limits of 0.1 to 0.6 MeV are obtained for proton targets, and the $J^P = 1/2^+$ assignment.  For neutron targets limits from 0.26 and 1.7 MeV are obtained. Much smaller limits of $< 0.1$ MeV are obtained for $J^P = 3/2^-$, while for $J^P = 1/2^-$, limits of 1 to 2.5 MeV are extracted. Although these limits are model-dependent, taken together they still present formidable constraints on the $\Theta^+$ width. Hadronically decaying resonances with total decay widths of less than a few MeV would seem unusual, but widths of a few hundred keV or less would make the existence of the state highly unlikely. In order to have quantitative tests of the LEPS results, which is the only remaining low energy photoproduction experiment with a positive signal, the old and new results from LEPS should be turned into normalized cross sections and compared to the CLAS data on deuterium.        

When the dust will have settled on the issue of narrow pentaquark baryons, we will have learned a lot about the physics of hadrons, no matter what the final outcome will be.

\section*{Acknowledgment}

I wish to thank Takashi Nakano for providing me with unpublished results from the LEPS experiment.

\end{document}